# High-harmonic generation from a subwavelength dielectric resonator

Anastasiia Zalogina[1,2], Luca Carletti[3], Anton Rudenko[4], Jerome V. Moloney[4], Aditya Tripathi[1],

Hoo-Cheol Lee[5], Ilya Shadrivov[1], Hong-Gyu Park[5], Yuri Kivshar[1], and Sergey S. Kruk[1]

[1]Nonlinear Physics Centre, Research School of Physics, The Australian National University, Canberra ACT 2601, Australia
[2]Research School of Biological Sciences, The University of Adelaide, Adelaide SA 5005, Australia
[3]University of Brescia, Brescia 25123, Italy
[4]Arizona Centre for Mathematical Sciences and Wyant College of Optical Sciences, University of Arizona, Tucson, Arizona 85721, USA
[5]Department of Physics, Korea University, Seoul 02841, Republic of Korea.

**Abstract**

Higher-order optical harmonics entered the realm of nanostructured solids being observed recently in optical gratings and metasurfaces with a subwavelength thickness. Structuring materials at the subwavelength scale allows for resonant enhancing of the efficiency of nonlinear processes and reducing the size of high-harmonic sources. Here we report the observation of up to a seventh harmonic generated from a single subwavelength resonator made of AlGaAs material. This process is enabled by careful engineering of the resonator geometry for supporting optical modes associated with a quasi-bound state in the continuum in the mid-infrared spectral range at around $\lambda=3.7$ μm pump wavelength. The resonator volume measures $\sim 0.1\ \lambda^3$. The resonant modes are excited with an azimuthally polarized tightly focused beam. We evaluate the contributions of perturbative and non-perturbative nonlinearities to the harmonic generation process. Our work proves the possibility to miniaturize solid-state sources of high harmonics to the subwavelength volumes.

# INTRODUCTION

Subwavelength particles made of high-index dielectrics can be shaped to support resonant optical modes capable of enhancing the efficiency of nonlinear light-matter interaction by orders of magnitude [1]–[3]. Nonlinear optics of subwavelength resonators has originally been dominated by the studies of the second- and third-order nonlinear processes such as $2^{nd}$ and $3^{rd}$ harmonic generation [4], [5]. Such lower-order nonlinear optical processes are conventionally described within a perturbative approximation which assumes that nonlinear material polarization is only a small perturbation to its linear counterpart, and thus an incident light beam can only slightly perturb materials' properties. However, sufficiently intense excitations with pulsed laser systems can bring materials into the regime of non-perturbative nonlinearities. A canonical process of nonperturbative nonlinear optics is high-harmonics generation (HHG) [6], [7].

Historically, HHG has been associated with nonlinearities of gases and plasma [8], [9]. Only recently HHG from solids has been demonstrated [10]–[13] opening a pathway to higher-order nonlinearities in nanophotonics. HHG has been observed in two-dimensional layouts of subwavelength elements, such as metasurfaces [14]–[21]. However, the sizes of such structures remained large in two lateral dimensions. Resonant enhancement of HHG in the demonstrated layouts relied on resonances inherently limited to extended systems, such as electromagnetically-induced transparency (EIT) [15], or symmetry-protected bound states in the continuum (BICs) [16]. The reliance on collective modes of extended systems hindered to this date the miniaturization of a HHG source to the subwavelength scale in all three dimensions.

In what follows, we demonstrate a HHG source scaled down to a subwavelength volume of a single dielectric nanoparticle (see Figure 1) measuring 0.4 pump wavelength in height and 0.55 wavelength in diameter. We see the $5^{th}$ and $7^{th}$ optical harmonics in the visible spectral range generated from a resonator excited with a mid-infrared pulsed laser at around 3.5-4 μm. The observation is enabled by the enhancement of local fields via a resonant mode supported by a stand-alone particle. In contrast to collective resonances of 2D systems, the resonant response emerges as an interference effect between localized modes resembling the Friedrich-Wintgen scenario [22] of the bound states in the continuum (BICs) [23].

BICs were first proposed in quantum mechanics as localized electron waves with energies embedded within the continuous spectrum of propagating waves [24]. Since then, BICs have attracted interest in other branches of physics, including photonics where they manifested themselves as resonances with large quality factors (Q factors) limited only by finite sample size, material absorption, and structural imperfections. Optical BICs have first been studied in extended systems [25], [26]. More recently, the concept of BICs enabled a pathway towards high-Q factor modes in individual, stand-alone subwavelength dielectric resonators [27]. We note that the Q-factor of a realistic subwavelength resonator stays finite, and therefore the modes associated with the bound states in the continuum are known as quasi-BICs. Single nanoparticles hosting such modes have demonstrated enhancement of second-harmonic generation [4], as well as multiphoton luminescence [28]. Recent theoretical predictions suggested the enhancement of both direct and cascaded generation of fifth harmonic in quasi-BIC nanoparticles [29].

# RESULTS

We employ the concept of quasi-BICs for a mid-infrared AlGaAs resonator to generate HHGs in the visible spectral range. The quasi-BIC resonance in our design appears as a dark mode for a linearly polarized Gaussian beam. To couple to the quasi-BIC state, we use a structured light excitation with an azimuthally polarized mid-infrared beam. Strong mid-infrared excitation drives the resonator beyond the boundaries of perturbative nonlinear optics.

To get a theoretical insight into the resonant behavior under strong laser excitation, we introduce a self-consistent model which provides information about the dynamics of spatial inhomogeneous electron-hole density distribution inside an isolated resonator, and its nonlinear response which yields higher-order harmonic spectra. Simulation results support the experimental findings predicting the enhancement of the harmonics by several orders of magnitude compared to an unstructured sample, non-resonant geometry, or off-resonant laser irradiation condition and confirming their nonperturbative origins.

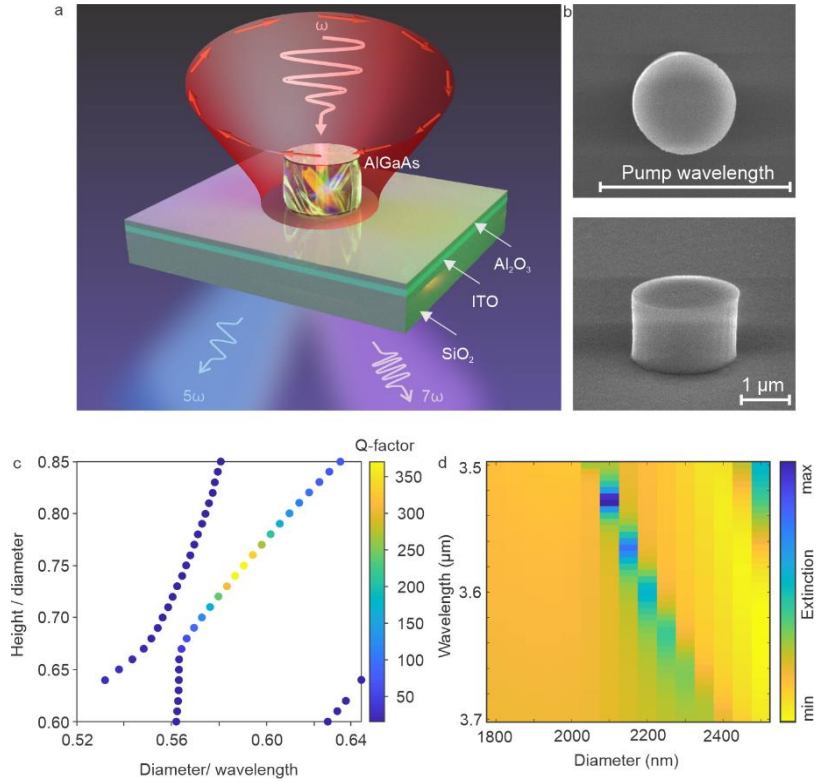

**Figure 1. Subwavelength resonator for a high-harmonic generation.** (a) Schematics: the light of a frequency ω is incident on a resonator which is placed on a substrate, which excites high harmonics. (b) Scanning electron microscope images of the fabricated resonator. (c) Dependence of the resonator's modes on geometrical parameters and wavelength. Avoided crossing of two modes leads to the enhancement of the Q-factor (indicated with the color scheme). (d) Extinction of the incident light at around the resonant wavelength and resonator's diameter.

**Linear calculations**

We design the disk resonator from AlGaAs material with [100] crystal axis orientation. The resonator is placed on a substrate with a buried indium tin oxide (ITO) layer (see Figure 1a). The ITO has zero real part of the permittivity at around 1250 nm wavelength. Thus, in the spectral range of the excitation and the lower harmonics wavelengths it behaves as a metal (with negative permittivity), acting as a back reflector and increasing the Q-factor of the resonant modes. However, it becomes a transparent dielectric (positive permittivity) in the spectral range of HHG. We find the optimal design of the subwavelength resonator using COMSOL eigenmode analysis (see details in Methods). Figure 1c shows the dependence of the resonator's Q-factor on the geometrical parameters (height-to-radius ratio) as well as on the wavelength of the incident light. Individual circles correspond to eigenmodes of the resonator. We see a dramatic enhancement of the Q-factor in the vicinity of an avoided crossing of the two resonator's modes. We additionally optimize the thickness of a spacer between the buried ITO layer and the resonator (see details in the Supplementary Materials). We next choose the resonator's radius and height such that the high-Q mode appears in the mid-infrared spectral range. Our calculations show the Q-factor approaching 350 for the resonator 1384 nm in height and 2050 nm in diameter, 700 nm thickness of aluminum oxide spacer, and 310 nm thickness of the

ITO. In Figure 1d we calculate the extinction of the resonator dependent on its geometrical parameters and wavelength in the vicinity of the high-Q mode shown in Figure 1c. Here we use an azimuthally polarized excitation beam which matches the polarization structure of the high-Q mode. We assume beam focusing with a numerical aperture of 0.56. We see a sharp local extremum of the extinction spectrum.

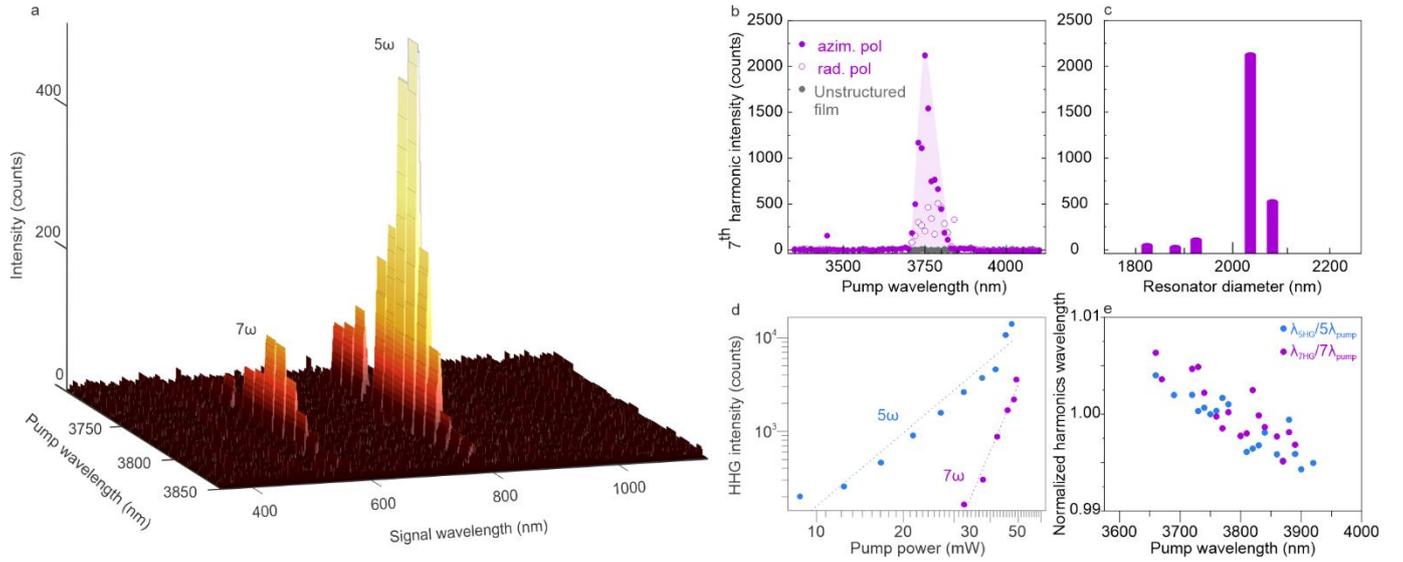

**Figure 2. Experimental observation of the 5$^{th}$ and 7$^{th}$ harmonics generation from a single resonator.** (a) Detected spectra vs fundamental wavelength. The resonant peaks correspond to the wavelengths of the 5$^{th}$ and 7$^{th}$ harmonics. (b) Normalized 7$^{th}$ harmonic intensity vs excitation wavelength. Filled dots – azimuthally polarized excitation. Hollow dots – orthogonally polarized excitation. Gray dots – signal from unstructured AlGaAs film with a thickness equal to the resonator's height. (c) 7$^{th}$ harmonic intensity at the resonant wavelength for several resonators' diameters. (d) Dependence of the power of the generated 5$^{th}$ and 7$^{th}$ optical harmonics on the pump power. Dashed lines are the fits with the dependences $P(5\omega) \sim P(\omega)^{2.6}$ and $P(7\omega) \sim P(\omega)^{3.5}$. (e) Frequency pulling of the harmonic wavelengths in the vicinity of BIC resonance.

**Fabrication**

Based on the theoretical design, we fabricate a set of standalone resonators from Al(0.2)Ga(0.8)As material with [100] orientation of the crystalline axis. We keep the height of the resonators fixed at 1384 nm, and we vary their radii. We use electron beam lithography followed by dry etching to define the geometry of the resonators. The resonators are then transferred to a glass substrate coated with 300 nm-indium tin oxide (ITO) and 700 nm-aluminum oxide layers.

**Optical Experiment**

We excite each resonator with 3300-4100 nm wavelengths with a tunable pulsed laser system and detect high-harmonic spectra in transmission (see details in Methods). Figure 2a shows experimental spectra of the 5th and 7th harmonics. We did not observe even-order high harmonics despite the non-centrosymmetric nature of the material of the resonator. Our theoretical calculations show (see Figure 3) that the nonperturbative regime of harmonics generation in our setting favors odd-order harmonics over even-order. The crystal orientation [100] of the sample further reduces the efficiency of higher-order even harmonics generation and collection. Our calculations suggest that in such resonators, the 4th harmonic intensity would be comparable to the 9th harmonic which is below our detection limit. We note that similar single subwavelength resonators were shown to produce enhanced multiphoton luminescence enhanced by Mie resonances when excited with linear polarization [28].

We next systematically study experimentally and theoretically the dependence of the 7$^{th}$ harmonic on the excitation wavelength. Figure 2b shows a pump wavelength scan for the resonator disk diameter corresponding to the BIC resonance. We see sharp resonant enhancement of the 7$^{th}$ harmonic signal at around 3750 nm excitation wavelength. We perform the same measurements for the orthogonal (radial) polarization of the incident beam and see the weaker signal of the 7$^{th}$ harmonic (see Fig. 2b). We note that from linear calculations only negligible HHG is expected for the radial excitation. We attribute a weaker, but detectable experimental signal for the radial excitation to experimental imperfections, such as imperfections of the mid-IR vortex retarder made in-house, aberrations of a single mid-IR lens used for focusing, slight deviations of the fabricated sample from perfect cylindrical shape, as well as relatively lower beam quality of the mid-IR output from the optical parametric amplifier.

We also assess the dependence of the HHG signal on the resonator size by scanning resonators with different diameters (see Figure 2c). We confirm the strong dependence of the performance of the resonators on the diameter. Our experimental observations of the HHG dependence of the excitation wavelength and the resonator size (Fig. 2b,c) are in good qualitative agreement with theoretical predictions of the extinction maximum (Fig. 1d).

For the optimal diameter and wavelength, we study the dependence of the intensities of the 5$^{th}$ and 7$^{th}$ optical harmonics on the pump power (Figure 2d). We derive the power scaling laws $P(5\omega) \sim P(\omega)^{2.6}$ and $P(7\omega) \sim P(\omega)^{3.5}$ for the 5$^{th}$ and the 7$^{th}$ harmonics respectively, that deviate significantly from the power dependence expected from the perturbative theory $P(5\omega) \sim P(\omega)^5$ and $P(7\omega) \sim P(\omega)^7$. We attribute the saturation in the harmonic yield power scaling to the non-perturbative nonlinearities, where at strong applied laser fields constant nonlinear susceptibilities of high orders $\chi^{(5)}(I), \chi^{(7)}(I)$ can no longer account for the nonlinear response. In this regime, the overall perturbative expansion of the electric susceptibility does not hold due to significant changes in the refractive index of the material under non-equilibrium ultrashort laser excitation. This saturation effect is common in bulk solids at intensities approaching TW/cm$^2$ [10]. We note that while in our experiments, the incident intensities are lower, the estimated intensities inside the nanodisk under resonant excitation conditions are in a similar range. We keep power below the laser damage threshold, which for our sample occurs at 50 mW for the resonant excitation.

We finally measure the wavelength of the fifth harmonic and the seventh harmonic versus the pump wavelength. For off-resonant excitation we detect slight deviations of the harmonic's wavelength from values $\lambda_{pump}/5$ and $\lambda_{pump}/7$ correspondingly. The BIC resonance pulls the spectral maxima of the HHG towards itself (see Fig. 2e); thus, the most efficient off-resonant generation occurs for intermediate pump wavelengths in-between the wavelength of the peak laser power and the wavelength of the BIC resonance. This effect resembles frequency pulling first studied in lasers [30].

**Nonlinear calculations**

Next, we theoretically study the nonlinear generation of optical harmonics. Upon intense ultrashort laser interaction, the nonlinear optical properties of materials are modified by photoexcitation of electron-hole carriers. This effect is particularly pronounced when high-intensity ultrafast light is confined in a nanoscale hot spot inside the subwavelength resonator. In this case, the yields of low- and higher-order harmonics in the emitted spectra are enhanced by being sensitive to laser-induced carriers. To investigate qualitatively the nonlinear propagation through a subwavelength resonator and harmonic generation, we apply a non-perturbative model based on full-vector nonlinear Maxwell equations supplemented by a nonlinear current, including the dynamic Drude response of laser-induced free carriers (see details in Methods).

Our numerical model allows us to apprehend three inter-coupled processes involved in the experiments:

- resonant field enhancement inside the resonator by considering (i) 3D geometry of the problem, (ii) specific laser irradiation conditions, and (iii) applied materials;
- photoexcitation of electron-hole pairs inducing changes in transient optical properties and 3D inhomogeneous distribution of electron plasma inside AlGaAs resonator;
- nonlinear transmission of the generated lower- and higher-order harmonics through the resonant system, following frequency mixing, non-perturbative enhancement, blue-shift by photo-excited carriers, and enhancement by ENZ thin film.

We note, however, that the description of photo-excitation processes is simplified in our calculations by considering only direct transitions to the conduction band and neglecting the contributions from inter-band polarization and indirect transitions [7]. This discrepancy can be further resolved by implementing the realistic multiple-band electronic structure and the quantum-based but more computationally demanding numerical methods.

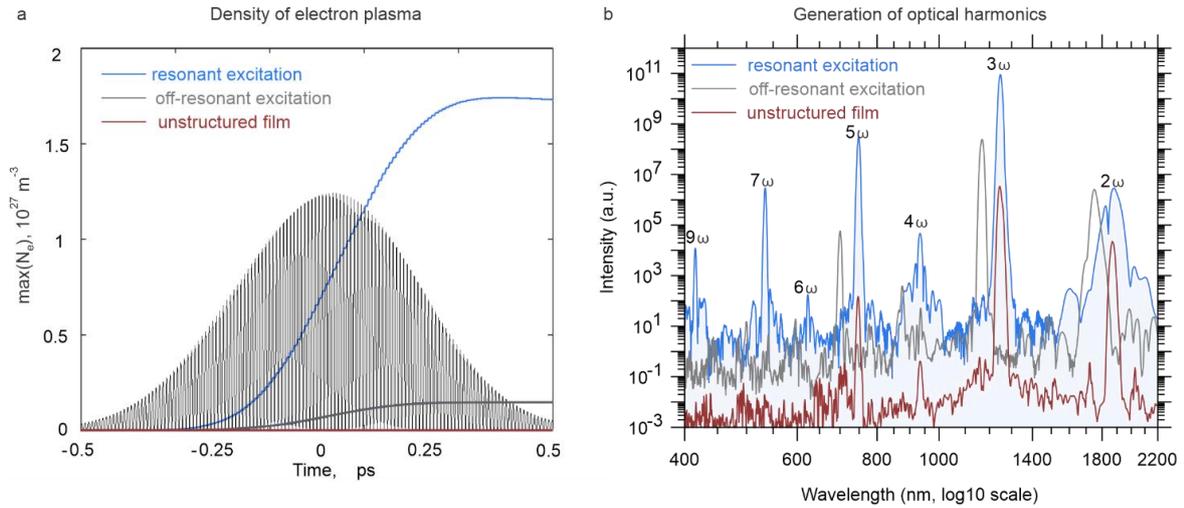

**Figure 3. Theoretical analysis of the high-harmonic generation.** (a) Temporal evolution of the maximum density of electron plasma inside subwavelength AlGaAs resonator leading to the non-perturbative nonlinear optical response. The black line visualizes the temporal profile of the excitation pulse (b) Theoretical transmission spectra of optical harmonics calculated by considering both the non-perturbative generation from the electron plasma and the perturbative cascaded generation from $\chi^{(2)}$ and $\chi^{(3)}$ nonlinear susceptibilities. (a,b) Blue: subwavelength AlGaAs disk 2050 nm in diameter, 1380 nm high, resonant wavelength excitation $\lambda = 3.75$ μm. Grey: subwavelength disk 2050 nm in diameter, 1380 nm high, off-resonance excitation with wavelength $\lambda = 3.5$ μm. Red: continuous AlGaAs film 1380 nm thick, wavelength of excitation $\lambda = 3.75$ μm.

The harmonic spectra depend strongly on the orientation of the crystal with respect to the laser polarization. Under normal incidence and crystal orientation [100], the odd harmonics show their maxima, whereas the even harmonics are unfavored and have not been observed when propagating along the crystal axis of bulk materials [10][31]. Nevertheless, the situation is different when a laser beam is tightly focused on a subwavelength structure. In this case, the second harmonic signal was detected in [5], while frequency mixing would also produce much weaker, but non-zero high-order even harmonics, which are not detected in the current experiment but obtained in numerical simulations.

We perform simulations of azimuthally polarized ultrashort laser pulse propagation through a subwavelength AlGaAs nanodisk, $Al_2O_3$, and ITO layers and analyze the resulting harmonic spectra in $SiO_2$ by taking the fast Fourier Transform (FFT) from the time history of the transmitted electric fields. The spectra consist of odd (3rd, 5th, 7th, 9th etc.) and less pronounced even (2nd, 4th, and 6th) harmonics as indicated in Figure 3. We point out that in these settings the 5th and 7th harmonics occur to be stronger than the 4th harmonic which remained undetectable in the experiment. We compare the spectra for unstructured AlGaAs and the optimal resonator of $D = 2050$ nm, both with the fixed AlGaAs thickness $H = 1380$ nm and for the incident power of 50 mW shown in Figure 3. The intensity of $2^{nd}$ and $3^{rd}$ harmonics are by two and four orders of magnitude smaller for the unstructured sample. The difference is even more pronounced for HHG, where the $7^{th}$ harmonic signal is strong and detectable for the resonator but below the noise floor for the unstructured sample (at least 9 orders of magnitude lower!). We note that the $7^{th}$ harmonic (~535 nm for $\lambda = 3.75$ μm excitation) occurs to be well above the bandgap of AlGaAs. We additionally perform numerical simulations for the nanodisks with off-resonant sizes as well as excited by off-resonant wavelength [see details in the Supplementary and an example in Figure S3b, the blue curve for an off-resonant $\lambda = 3.5$ μm]. We consider variations in the wavelength of the photo-ionization rates and permittivity of AlGaAs. We notice however that these effects are negligibly small compared to the stark difference in field enhancements inside the resonator. As a result, in off-resonant cases, much weaker carriers are generated (see Supplementary Materials), which strongly affects the yield of the generated HHG. The nonperturbative behavior of these harmonics is evidenced here. For resonant and off-resonant cases, there is a minor difference in the $2^{nd}$ harmonic, which has a mostly perturbative nature, related to $\overrightarrow{P^{(2)}}$, but the difference becomes significant already starting with the $3^{rd}$ harmonic. These simulation results agree well with the experimental measurements from Figure 2, indicating that it is problematic to detect higher-order signals if the resonator is off-resonant from the laser wavelengths. The comparison has been also done for different geometry by varying the diameter of the resonator from $D = 1.6$ μm to $D = 2.1$ μm (see Supplementary Materials). As expected, the off-resonant scenario results in less pronounced HHG and is thus undetectable experimentally. Simulations allow us to separate the contributions to harmonic orders provided by photo-excited carriers (nonperturbative origins) and by frequency mixing of $2^{nd}$ and $3^{rd}$ harmonics alone (perturbative origins). Furthermore, the power dependence of HHG is investigated for the resonant excitation by varying the laser power from 10 mW to 100 mW (see Supplementary Materials).

## DISCUSSION

We have observed the generation of the $5^{th}$ and $7^{th}$ optical harmonics from a single dielectric subwavelength resonator. The pronounced enhancement of the $7^{th}$ harmonic generation is driven by resonant modes associated with quasi-BIC states. Power dependences of the higher-order harmonics suggest the processes may include cascaded generation as well as nonperturbative regimes on nonlinear interactions. We have supported our findings by implementing a full-vector Maxwell-based approach with a nonlinear current, describing the response of photoexcited carriers. The calculated high-harmonic spectra are enhanced by several orders of magnitude compared to an unstructured sample, non-resonant geometry, or off-resonant laser irradiation condition. Our simulations suggest that the nanoscale confinement of electron plasma inside the subwavelength resonator is at the origin of pronounced nonperturbative odd harmonics. We have analyzed the contributions of the photo-excited carriers (nonperturbative origins) and cascaded frequency mixing of $2^{nd}$ and $3^{rd}$ harmonics alone (perturbative origins) and revealed that in our setting odd-order harmonics are favored over even-order harmonics. The generation of higher harmonics is one of the pathways toward light sources in the vacuum-UV and extreme-UV spectral ranges. Our results suggest how to miniaturize such light sources towards the subwavelength scale in solid-state systems by employing the physics of optical resonances in high-index dielectric particles.

## MATERIALS AND METHODS

### Linear Numerical Simulations

The numerical simulations for single subwavelength resonators were performed using COMSOL eigenmode analysis. The material permittivity values of all the materials are assigned at the designed wavelength of

3500 nm. The dispersion relation for the complex permittivity of ITO in the infrared is described by Drude-like dependence $\varepsilon_{ITO} = \epsilon_\infty - \left(\frac{\omega_p^2}{\omega^2} + i\omega\gamma\right)$, where $\epsilon_\infty = 4.068$, $\omega_p = 3.07 \cdot 10^{15}\ s^{-1}$, and $\gamma = 2.49 \cdot 10^{14}\ s^{-1}$ [29]. The Drude model agrees well with experimental ellipsometry data for our ITO films. Such dependence results in epsilon-near-zero wavelength (ENZ) $\lambda_{ENZ} = 1.25$ μm. While the ITO material is known to demonstrate strong nonlinear properties in the vicinity of the ENZ region [32], we note here that our excitation wavelengths in the mid-IR range are far from the ENZ wavelength positioned at around 1250 nm. The design for subwavelength resonators was defined for the parameters which reach the highest Q-factor with a height of 1384 nm, and a diameter of 2050 nm, which is placed on the top of an aluminum oxide spacer with a thickness of 700 nm, and the ITO layer with the thickness of 310 nm. The extinction of the resonator in Figure 1c is calculated using the azimuthal polarization and a numerical aperture of 0.56.

**Calculations of Nonperturbative Nonlinear Response**

To investigate qualitatively the nonlinear propagation through a subwavelength resonator and harmonic generation, we apply a non-perturbative model based on full-vector nonlinear Maxwell equations supplemented by a nonlinear current, including the dynamic Drude response of laser-induced free carriers as follows:

$$\begin{cases} \frac{\partial \vec{E}}{\partial t} = \frac{\nabla \times \vec{H}}{\varepsilon \varepsilon_0} - \frac{\vec{J_e}}{\varepsilon \varepsilon_0} - \frac{\partial}{\partial t}(\overrightarrow{P^{(2)}} + \overrightarrow{P^{(3)}}) \\ \frac{\partial \vec{H}}{\partial t} = -\frac{\nabla \times \vec{E}}{\mu_0} \qquad (1) \\ \frac{\partial \vec{J_e}}{\partial t} = -\nu_e \vec{J_e} + \frac{e^2 N_e}{m_e^*} \vec{E}, \end{cases}$$

where $\vec{E}$ and $\vec{H}$ are the electric and magnetic fields, $\varepsilon_0$ and $\mu_0$ are the vacuum permittivity and permeability, $\overrightarrow{P^{(2)}}$ and $\overrightarrow{P^{(3)}}$ are the second- and third-order perturbative polarizations, and $\vec{J_e}$, $N_e$, $\nu_e$, and $m_e^*$ stand for the time-dependent current density and carrier density, the collision frequency and the effective mass of laser-excited carriers respectively. Under the considered laser irradiation conditions, the carriers are excited only in the AlGaAs resonator, having a relatively small electron bandgap ($E_g = 1.67$ eV for 20% Al and 80% Ga content [33]), but not in ITO ($E_g \sim 4$ eV) and Al$_2$O$_3$/SiO$_2$ ($E_g \sim 9$ eV) substrates.

The system of equations (1) is solved by a finite-different time-domain (FDTD) method with an auxiliary differential equation (ADE) technique to introduce the nonlinear current $\vec{J_e}$ [34]. A fourth-order Runge-Kutta iteration method is applied to resolve the nonlinear equation for the electric fields $\vec{E}$. An azimuthally polarized pulse of $\theta = 522$ fs (FWHM) duration is focused with $NA = 0.56$ (beam waist radius of $w_0 \sim 5$ μm) on the surface of the resonator at $z = z_0$ as follows

$$E_{x,y}(x,y,t) = \exp\left[-4ln2\frac{(t-t_0)^2}{\theta^2} - \frac{x^2+y^2}{w_0^2}\right] \cos\left(i\omega t - ik\frac{x^2+y^2}{z_0}\right)[sin, cos]\left(arctg\frac{y}{x}\right), \quad (2)$$

where $t_0 = 2\theta$ is the time delay, $\omega = 2\pi c/\lambda$ is the frequency for the central laser wavelength $\lambda$, $c$ is the speed of light, and $k = 2\pi/\lambda$ is the wave vector. The pulse propagates along the $z$ axis, normal to the surface. In simulations, the considered structure includes an AlGaAs nanodisk of variable diameter $D$ and a fixed height of $H = 1380$ nm, an Al$_2$O$_3$ layer of 700 nm thickness, a 300 nm ITO film, and a SiO$_2$ substrate.

For the AlGaAs, the carrier density is evaluated by Keldysh photo-ionization rate $\frac{\partial N_e}{\partial t} = w_{PI}(|\vec{E}|, \lambda, E_g, m_e^*) - N_e/\tau_{rec}$ for the electric field $\vec{E}$ and laser wavelength $\lambda$, whereas constant $E_g = 1.67$ eV, $m_e^* = 0.07\ m_e$, $\tau_{rec} = 11$ ps, and $\nu_e \sim 10^{13}\ s^{-1}$ are adopted from Ref. [35] in the ionization model. In the absence of carriers, the third-order Kerr-like response is given by polarization $\vec{P^{(3)}} = \varepsilon_0 \chi^{(3)}(\vec{E}\vec{E})\vec{E}$ with $\chi^{(3)} = 3.4 \cdot 10^{-19} m^2/V^2$ [36] and the second-order response for non-centrosymmetric material by $\vec{P^{(2)}} = \varepsilon_0 \chi^{(2)}(E_y E_z, E_x E_z, E_x E_y)$ [37] with $\chi^{(2)} \sim 300$ pm/V [38]. We neglect here the contribution of surface second-order nonlinearities.

The optical properties for ITO are included by the ADE technique to FDTD Maxwell solver [34], considering the dispersion relation with $\lambda_{ENZ} = 1.25$ µm from Ref. [28]. For the central wavelength $\lambda = 3.75$ µm corresponding to the resonant excitation of the nanodisk, the other non-excited materials are transparent and their permittivities are given by $\varepsilon_{AlGaAs} \sim 3.1^2$, $\varepsilon_{Al2O3} \sim 1.7^2$, and $\varepsilon_{SiO2} \sim 1.4^2$.

**Resonators Fabrication**

The resonators were fabricated from Al(0.2)Ga(0.8)As material with [100] orientation of the crystalline axis with the method of epitaxial film growth. Next, a PMMA mask is defined on an AlGaAs wafer using electron-beam lithography. The vertical pillar structure is fabricated using chemically assisted ion-beam etching with a mixture gas of Ar and $Cl_2$, and the PMMA mask is removed using $O_2$ plasma. The AlInP sacrificial layer is wet etched with a dilute $HCl/H_2O$ (3:1) solution. Then, the AlGaAs nanodisk is picked up using the polypropylene carbonate (PPC)-coated polydimethylsiloxane (PDMS) stamping method. Finally, the PPC is separated from the PDMS by applying heat to the PDMS, to transfer the nanodisks to a glass substrate with 300 nm indium tin oxide (ITO) and 700 nm aluminum oxide layers. The detailed fabrication procedure is shown in Figure S7.

**Optical Measurements**

We excite the resonators with 3300-4100 nm wavelengths with a tunable pulsed laser system (Ekspla Femtolux femtosecond laser and MIROPA Hotlight Systems optical parametric amplifier, output 522 fs, 5.14 MHz repetition rate). The output power is attenuated by a pair of mid-infrared wire-grid polarizers to a level not exceeding 50 mW (average power). The linear polarization of the laser beam is converted into an azimuthal polarization by a silicon metasurface vortex retarder fabricated in-house [39]. The mid-infrared radiation is focused with an aspheric lens with 0.56 NA. The laser beam diameter is adjusted (widened by a telescope made of a pair of achromatic doublet lenses) to fit the diameter of the aspheric lens. The beam waist radius $w_0$ of the focused azimuthal beam is assumed to be ~20% larger than the diffraction-limited spot of a Gaussian beam [40], e.g. 0.73 λ/NA ~ 5µm at around the resonant wavelength. We thus estimate the incident power density in the focal spot at the levels of $10^{10}$ W/cm². The sample illumination with the mid-infrared beam was monitored on a camera Tachyon 16 (NIT) in reflection configuration with the use of a 50/50 beam splitter. The power of the incident beam was monitored with a Thorlabs power meter S405C. The nonlinearly generated light is collected in transmission with an objective Mitutoyo ×100 0.7 NA (achromatic spectral range 400-1800 nm). The light is detected with a Peltier-cooled CCD camera Starlight Xpress with an f=150mm achromatic doublet lens used as a camera objective. The spectra are detected with a spectrometer Ocean Optics QEPro. The spectrometer uses a silicon CCD array with a detection range of up to 1.1 µm. It is therefore suitable to detect optical harmonics starting from the 4th order and higher. In our experiments we leave out consideration of the second and third harmonic which were studied in detail in the past in similar systems [5], [33]. We note that the 2nd and 3rd harmonic are undetectable in transmission configuration as the buried ITO layer acts as a metallic mirror at their wavelengths.

**SUPPLEMENTARY MATERIALS**

Figure S1. Theoretical calculations of nonlinear intensity and maximum electron density distributions in subwavelength resonator at 3.75 µm.
Figure S2. Theoretical calculations of nonlinear intensity and maximum electron density distributions in subwavelength resonator at 4 μm.

Figure S3. Theoretical calculations of high harmonic spectra for different laser powers and resonant excitation conditions.

Figure S4. Theoretical calculations of high harmonic spectra for resonant and off-resonant excitation conditions and for different diameters and excitation conditions.

Figure S5. Theoretical calculations of high harmonic spectra in case of subwavelength resonator and unstructured sample and comparison of the spectra produced by both perturbative and non-perturbative nonlinearities.

Figure S6. Theoretical calculations of the Q-factor as a function of the coating layer thickness.

Figure S7. Schematics of the fabrication process.

**Acknowledgements.** Y.K. thanks David Reis for highly insightful and stimulating discussions of the results. **Funding.** Australian Research Council (DE210100679, DP210101292), Samsung Research Funding and Incubation Centre of Samsung Electronics (SRFC-MA2001-01), Air Force Office for Scientific Research (FA9550-19-1-0032 and FA9550-21-1-0463), the International Technology Centre Indo-Pacific (ITC IPAC) via Army Research Office (FA520921P0034), and the National Research Foundation of Korea (NRF) funded by the Korean government (2021R1A2C3006781). **Author contributions:** A.Z., Y.K. and S.K.

conceived this research project; H.-C. L. and H.-G. P. fabricated resonators; A.Z., S.K. and A.T. carried out optical measurements; L.C. carried out linear numerical simulations and mode analysis; A.R. and J.V.M. carried out nonlinear numerical simulations; all the authors analysed the data and contributed to writing the manuscript. **Competing interests:** The authors declare that they have no conflict of interests. **Data and materials availability:** All data needed to evaluate the conclusions in the paper are present in the paper and/or the Supplementary Materials.